\documentclass[aps,pra,onecolumn,preprint,showpacs,groupedaddress,superscriptaddress,nofootinbib,floatfix,preprintnumbers,longbibliography]{revtex4-2}
\usepackage{graphicx}
\usepackage{dcolumn}
\usepackage{appendix}
\usepackage{bm}

\usepackage[usenames,dvipsnames]{xcolor}
\usepackage{subfigure}
\usepackage{bbm}
\usepackage{enumerate}
\usepackage[
  bookmarks=true,
  colorlinks,
  linkcolor=black,
  urlcolor=black,
  citecolor=black,
  plainpages=false,
  pdfpagelabels,
  final,
  breaklinks=true
]{hyperref}

\usepackage{titlesec}
\usepackage{array}
\usepackage{dsfont}
\usepackage{physics}

\begin{document}


\title{High-Order Harmonic Generation Driven by Perfect Optical Vortex Beams: Exploring the Orbital Angular Momentum Upscaling Law}

\author{Bikash Kumar Das}
\email{bikash.das@campus.technion.ac.il}
\affiliation{Department of Physics, Guangdong Technion - Israel Institute of Technology, 241 Daxue Road, Shantou, Guangdong, China, 515063}
\affiliation{Department of Physics and Solid State Institute, Technion -- Israel Institute of Technology, Haifa, 32000, Israel}
\affiliation{Guangdong Provincial Key Laboratory of Materials and Technologies for Energy Conversion, Guangdong Technion - Israel Institute of Technology, 241 Daxue Road, Shantou, Guangdong, China, 515063}

\author{C. Granados}
\affiliation{Department of Physics, Guangdong Technion - Israel Institute of Technology, 241 Daxue Road, Shantou, Guangdong, China, 515063}
\affiliation{Guangdong Provincial Key Laboratory of Materials and Technologies for Energy Conversion, Guangdong Technion - Israel Institute of Technology, 241 Daxue Road, Shantou, Guangdong, China, 515063}
\affiliation{Technion - Israel Institute of Technology, Haifa, 32000, Israel}

\author{M. Krüger}
\affiliation{Department of Physics and Solid State Institute, Technion -- Israel Institute of Technology, Haifa, 32000, Israel}

\author{Marcelo F. Ciappina}
\email{marcelo.ciappina@gtiit.edu.cn}
\affiliation{Department of Physics, Guangdong Technion - Israel Institute of Technology, 241 Daxue Road, Shantou, Guangdong, China, 515063}
\affiliation{Guangdong Provincial Key Laboratory of Materials and Technologies for Energy Conversion, Guangdong Technion - Israel Institute of Technology, 241 Daxue Road, Shantou, Guangdong, China, 515063}
\affiliation{Technion - Israel Institute of Technology, Haifa, 32000, Israel}

\date{\today}

\begin{abstract}

Orbital angular momentum (OAM) light beams for high-order harmonic generation (HHG) provide an additional degree of freedom to study the light-matter interaction at ultrafast timescales. A more sophisticated configuration is a perfect optical vortex (POV) beam, a light beam with a helical wavefront characterized by a phase singularity at its center and an azimuthal phase variation. POV beams are characterized by a radial profile which is independent of the OAM. Here we study the non-perturbative process of gas-phase HHG using a linearly polarized POV beam. We observe that the harmonics are emitted with similar divergence due the perfectness of the POV-driven harmonics. Furthermore, the topological charge upscaling is rigorously followed. We show that a POV beam is more advantageous than that of the Laguerre-Gaussian beam for cases where a large topological charge with a small core size is required. Our research establishes a pathway for producing bright structured extreme ultraviolet (XUV) coherent radiation sources--a pivotal tool with multifaceted applications across various technological domains.



\end{abstract}

\maketitle


\section{Introduction.}

Attosecond science delves into the ultrashort timescales, offering unprecedented insights into ultrafast processes that govern the behavior of light and electrons in matter~\cite{Krausz2009,Krausz2014,Calegari2016}. At the heart of attosecond science is the generation of attosecond light pulses~\cite{Paul2001,Hentschel2001} in the extreme ultraviolet (XUV) regime with record-short pulses down to 43\,as~\cite{Gaumnitz2017}. Such temporally tailored light pulses serve as the primary tool for probing and controlling and observing rapid events in atoms, molecules, solids and liquids at their most fundamental level.


In the spatial domain, it is a well-known fact that light beams carry both spin (SAM) and orbital angular momentum (OAM)~\cite{bliokh2015spin}. The SAM stems from the polarization of the light field whereas the OAM is related to its spatial distribution and the formation of a helical phase front. Allen et al.~\cite{allen1992orbital} revolutionized the field of singular optics in 1992 through their study on Laguerre-Gaussian (LG) beams, which are solutions to the paraxial Helmholtz equation in cylindrical coordinates. These beams are characterized by an annular intensity distribution with a twisted phase. Light beams carrying OAM, such as LG beams and Bessel-Gauss (BG) beams, have been employed in a multitude of novel applications such as in particle trapping and manipulation~\cite{padgett2011tweezers}, microscopy~\cite{fürhapter2005spiral}, free space optical communication~\cite{wang2012terabit}, light-matter interaction in atoms and molecules~\cite{scholz2014absorption,paufler2018strong},  recognition of chiral molecules~\cite{forbes2019raman,ward2016resolving}, for instance. Structured light beams carrying OAM also open up new perspectives for attosecond science as they enables studies of strong-field light-matter interactions, such as high harmonic generation (HHG)~\cite{Ferray1988,McPherson1987}, the hallmark effect in the generation of attosecond light pulses. In HHG, a strong laser field interacts with an atomic target and generates new frequencies (or harmonics) that are not contained in the driving laser field. Thus, HHG is a highly non-linear and non-perturbative process, widely used as a tabletop source of coherent  XUV and soft X-ray (SXR) radiation. A pioneering experiment by Z\"urch et al.~\cite{zurch2012strong} studied HHG in atomic gases using LG pump beams. Their findings revealed that the OAM of different harmonic orders is the same as the OAM of the fundamental laser field. However, this contradicts the conservation of OAM, as predicted from the Lewenstein model of HHG~\cite{maciejSFA,symphony}. They attributed this deviation to the propagation of the harmonic fields from the interaction region to the detector. Later, the upscaling law of OAM or topological charge (TC) was verified by Garcia et al.~\cite{garcia2013attosecond}. Here, the authors numerically simulated the propagation of harmonic fields at each point of the target medium towards the detector. These theoretical findings were later experimentally verified by Gariepy et al.~\cite{gariepy2014creating}. Since then, many experimental studies have been carried out focusing on, for instance, generating helical electron pulses~\cite{geneaux2016synthesis} or creating time-varying OAM (self-torque of light beam) in HHG~\cite{laura2019generation}. Though it is relevant to study the HHG driven by LG beams, there exist some difficulties associated with them, namely (i) the scaling of the radius of maximum intensity and beam size with the fundamental OAM--this restricts the nominal intensity required for HHG and (ii) the characterization of the harmonic vortices. Here, their short wavelength leads to a low fringe spacing in the interferograms which becomes difficult to resolve. These issues can be addressed with a more sophisticated beam profile, popularly known as the perfect optical vortex (POV) beam. It was shown by Ostrovsky et al.~\cite{ostrovsky2013generation} that the POV beams have an OAM-independent core size and intensity distribution. Furthermore, a POV beam can also be generated via the optical Fourier transform of a BG beam~\cite{vaity2015spatial} or using planar Pancharatnam-Berry phase elements~\cite{Liu2017}. 

In this contribution, we study the non-perturbative process of HHG in atomic gases using a fundamental POV beam of OAM, $l=1$. We employ the thin slab model (TSM) to calculate the near-field amplitude and phase profile of different harmonic orders when the slab, which resembles the atomic gas target, is placed at the focus of the beam. Then, we use the Fraunhofer diffraction integral to calculate the far-field intensity and phase profile of those harmonics.
In particular, we show that the divergence of different harmonic orders is the same irrespective of their OAM, and the OAM upscaling law is strictly followed. 

\section{Proposed experimental setup and description of the POV beam}

\begin{figure}[h!]
\includegraphics[width=.8\textwidth]{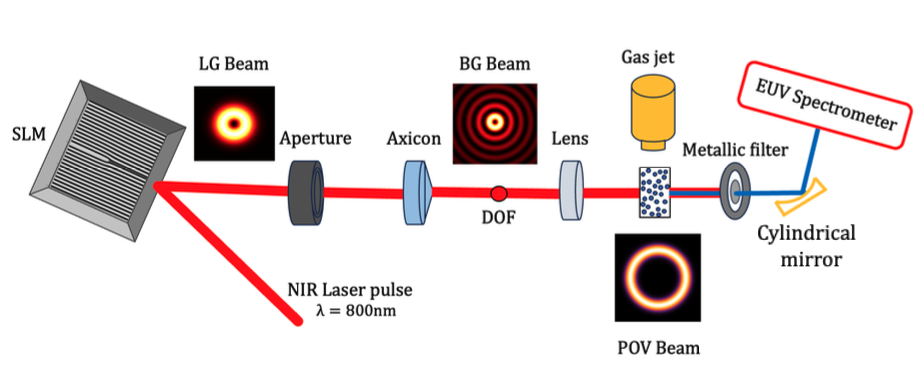}
\caption{\label{ExpSet}Proposed experimental set up to generate gas-phase HHG using a POV beam (SLM-Spatial Light Modulator; DOF-Depth of Focus).}
\end{figure}

The experimental realization of the POV driven HHG in atomic gases has not been explored yet. In the following, we would like to propose an experimental setup that could be implemented to test our theoretical model. The proposed experimental set up to investigate gas-phase HHG using a spatially structured, linearly polarized intense POV beam is shown in Fig.~\ref{ExpSet}. A phase mask, consisting of the sum of a spiral and an axicon phase (with an axicon parameter $a=44.7$ mm$^{-1}$), is loaded onto a liquid crystal-spatial light modulator (LC-SLM) operated in phase-only mode. An ultrashort ($\tau_{p}=61.4$ fs) intense NIR ($\lambda=800$ nm) Gaussian-shaped pulse ($\omega_0=150$~$\mu$m) is incident on the SLM, which in turn, generates a BG beam ($\omega_g=509$~$\mu$m) of topological charge (TC) $1$. Then, a POV beam of TC $1$ ($R=1.71$ mm) is generated at the far-field i.e., at the focus of the lens, when an optical Fourier transform of the BG beam is performed via a plano-convex lens ($f=300$ mm).  The laser beam, and optics parameters are chosen in such a way that the peak intensity of the POV beam at the focus is $1.57 \times 10^{14}$ W/cm$^2$. An atomic gas jet is positioned exactly at the beam focus position. The interaction of the intense POV beam with the atomic gas results in the generation of XUV harmonics. Later, a metallic filter (such as a thin Al filter) is used to block the fundamental POV beam radiation. As a result, the XUV harmonics pass through the filter, and are focused by a cylindrical mirror onto the XUV spectrometer. The XUV spectrometer consists of a grating, a micro channel plate, and a CCD camera, where the structured high-order harmonics are spectrally characterized.

\begin{figure}[h!]
\includegraphics[width=0.7\textwidth]{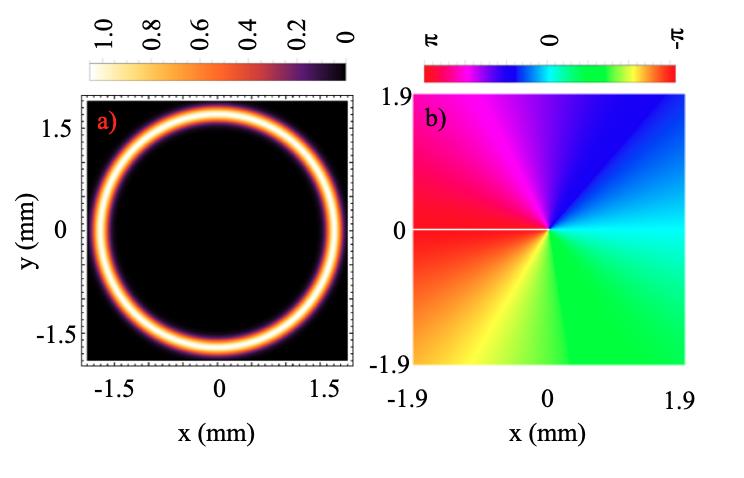}
\caption{\label{InteProf} Transverse intensity (a) and phase profile (b) of a POV beam carrying a TC $l=1$.}
\end{figure}

Our theoretical analysis on HHG using a POV pump beam starts by defining the spatial complex field amplitude at the source plane where the gas target is placed in our model (see Fig.~\ref{ExpSet}). Thus, we can write 
\begin{equation}
    E(r,\theta,z=0)=E_0e^{-\frac{r^2+R^2}{\omega_0^2}}e^{il\theta}I_l\left( \frac{2rR}{\omega^2_0} \right),
\end{equation}
where $E_0$, $R$, $l$ , and $\omega_0$ represent the peak field amplitude, ring radius, topological charge (interchangeably OAM), and half-ring width of the beam respectively. Here, $I_l$ is the modified Bessel function of the first kind of order $l$ (see Appendix \ref{A1} for more details). The transverse intensity and phase profile of a POV beam of TC $1$ are shown in Figs.~\ref{InteProf}(a) and~\ref{InteProf}(b), respectively. The transverse intensity profile and the  size of the dark core of the POV beam are independent of the TC (see Fig.~4 in Appendix \ref{A1}). Furthermore, unlike LG beams, there is only a single bright ring in the intensity profile. If the ring radius of the beam is much larger than its ring width, the modified Bessel function of the first kind can be approximated as $I(2rR/\omega^2_0) \approx e^{2rR/\omega^2_0}$ for $R\gg \omega_0$. In this way, the complex field amplitude of the POV beam becomes $E(r,\theta,z=0)=E_0 e^{-(r-R)^2/\omega_0^2}\times e^{il\theta}$. The OAM, $l$, induces an azimuthal phase distribution, i.e., the phase curls around the axis of singularity marked as a white line in Fig.~\ref{InteProf}(b). The modulus square of the complex field amplitude is independent of the azimuthal angle $\theta$, but depends on $r$, which leads to the formation of a bright ring at $r=R$ in the transverse intensity profile. Hence, the radius of maximum intensity defines the divergence of the POV beams. We found that POV beams of higher TCs show similar divergence. This is in contrary to LG beams, where the divergence, and hence the beam size, are tightly linked to the TC. This, in general, poses a limit to the use of high-order spatially structured light beams, such as LG and BG beams, for HHG. This is because an increment in the TC leads to an enlargement of the beam size, that, in turn, reduces the peak intensity at the focus. In certain scenarios, this intensity may fall below the threshold intensity required for gas-phase HHG.

\section{Microscopic and macroscopic aspects of gas-phase HHG} 

When an intense laser field ($I\approx10^{13}-10^{14}$ W/cm$^2$) interacts with an atomic gas target, new frequencies, known as high-order harmonics, are generated as a consequence of the strong nonlinear response of the target medium. Experimentally, the detected harmonics represent the macroscopic response of a large collection of single atomic emitters located in the interaction volume. Thus, the HHG process should be theoretically modeled in two stages, namely: (i) the single-atom response to the strong driving field and (ii) the phase-matched coherent summation of those single atom responses. Considering that the strong driving field propagates through an ionized medium, propagation effects should typically be taken into account. For calculating the single atom response, it is customary to solve the time-dependent Schrödinger equation (TDSE) under the single active electron (SAE) and dipole approximations (see the Appendix \ref{A2}). However, the TDSE requires large computational resources and is unable to provide information about the underlying physical mechanisms. To circumvent these problems, the strong-field approximation (SFA)~\cite{maciejSFA}, which is the quantum mechanical counterpart of the semi-classical three-step model~\cite{corkum1993plasma,Kulander1993}, can be employed. The SFA allows for the full solution of the electron dynamics if one considers that (i) the strong laser field does not couple with any bound state beyond the ground state;
(ii) the ground state depletion is small 
and (iii) there is no influence of the Coulomb potential on the ionized electron in the continuum.
The dipole approximation holds true in our problem, as the tunneled electron's classical excursion distance is of the order of 1 a.u.(0.53 \AA), which is extremely small compared to the fundamental POV beam wavelength ($\lambda=800$ nm). Thus, the electron does not feel the spatial structure of the driving field and we can use a linearly polarized field in our single-atom simulations.

To calculate the propagation effects in the fundamental and harmonic fields, Maxwell's wave equations and the Fraunhofer diffraction integral in the near and far-field, respectively, are usually employed. However, this method is computationally very demanding and requires complete knowledge of the various factors contributing to the phase matching of the fundamental and harmonic fields. A simpler and more efficient approach for calculating the near and far-field amplitudes and phase profile of different harmonic orders is the thin slab model (TSM). In the TSM, a thin 2D slab (in which atoms are placed randomly) is positioned perpendicular to the fundamental beam propagation axis. Consequently, this model only considers the transverse phase matching and avoids the longitudinal one. It is justified by the fact that the slab used in our model is extremely thin. Hence, the effects of longitudinal phase matching can be safely disregarded. The TSM, combined with SFA, allows for the calculation of the near-field amplitude and phase profile of different harmonic orders.

In our case, we utilize the TSM to investigate the interaction of gas atoms with an intense spatially structured POV beam carrying an OAM of $l=1$. By doing so, we gain insight into the physical origin of the structure of the XUV vortices generated via HHG. We assume that the fundamental POV field remains undistorted throughout the interaction with the gas atoms and during propagation along the medium. This assumption is justified by the thinness of the medium and the relatively low gas pressure, which in turn makes the amount of ionization very small. This reduces the complexity of propagating both the fundamental and harmonic fields from the target medium to the detector.


\section{Law of OAM upscaling for vortex harmonics}
In the TSM presented in our work, we position the slab exactly at the beam focus, i.e., at $z=0$, and employ the SFA to compute the single atom responses. This approach avoids the phase contributions arising from the curved wavefront of the beam profile (focal phase) and beam focusing (Gouy phase), simplifying our calculations. Subsequently, the slab can be displaced to any value of $z$ with respect to the beam focus position, and the same computational procedure can be repeated to determine the near-field amplitude and phase profile of different harmonic orders. Thus, we can simulate a true 3D target interacting with an intense laser field for HHG. The effectiveness of this model can be verified by the fact that the predictions for the divergence profile of different harmonic orders match exactly the calculations coming from the conventional model, which propagates both the fundamental and harmonic fields using Maxwell’s equations.
The contribution of the $j^{\mathrm{th}}$ quantum path (short or long) to the $q^{\mathrm{th}}$ harmonic emitted at a slab placed at $z=0$ is given by:
\begin{equation}
\label{near}
    A_q^{j(\text{near})}(r',\theta ')=B(j)\abs{U(r')}^pe^{iql\theta ' +i\alpha _q^j\abs{U(r')}^2}, 
\end{equation}
where, $B(j)=(C/\tau^j)^{3/2}$ and $p$ represents the scaling factor, which must be extracted for different harmonic orders from the SFA spectra (see Appendix \ref{A2}). The amplitude terms of the fundamental POV beam are grouped in $U(r')$. Here, $\tau^j$ denotes the excursion time associated with the $j^{\mathrm{th}}$ quantum path (long or short), which affects the HHG conversion efficiency and is related to the quantum diffusion of the electronic wavepacket, $\alpha_q^j$ is a quantum path-dependent strong-field parameter that relates the dipole phase to the fundamental field intensity~\cite{maciejphase}, and $(r',\theta')$ represent the near-field polar coordinates. Generally, the $q^{\mathrm{th}}$ harmonic order amplitude, $U_q$, scales differently with the fundamental field amplitude, $U_f$, in perturbative and non-perturbative regimes, i.e., $U_q\propto U_{f}^p$, with $p=q$ ($p<q$) in the perturbative (non-perturbative) regime. From Eq.~(\ref{near}), it is evident that extracting the value of $p$ is crucial for calculating the harmonic field amplitude at the near-field. To obtain this scaling for three different harmonic orders (19$^{\mathrm{th}}$, 21$^{\mathrm{st}}$, and 23$^{\mathrm{rd}}$ in our numerical simulation), we utilize three different intensities of the input field and compute the harmonic amplitudes using the SFA (as shown in the Appendix \ref{A2}). Subsequently, we extract the value of $p$ for different harmonic orders. 
It is important to note that the values of $p$ obtained in our simulation ($p_{19}=3.9$, $p_{21}=3.4$, and $p_{23}=3.6$) are very similar, as these harmonic orders fall in the plateau region of the HHG spectrum, where the harmonic yield is almost constant. Furthermore, we employ the Fraunhofer diffraction formula to compute the far-field amplitude and phase profiles of these harmonics, i.e.
\begin{eqnarray}
A_q^{j(\text{far})}(\beta, \theta)&\propto& B(j)e^{iql\theta}i^{ql} 
 \\ &\times&  \int_0^\infty r' dr' |U(r')|^pJ_{ql}\Bigg( \frac{2\pi\beta r'}{\lambda_q} \Bigg)e^{i\alpha_q^j \abs{U(r')}^2},\nonumber
\end{eqnarray}
where $\lambda_{q}=\frac{\lambda}{q}$ is the wavelength of the $q^{\mathrm{th}}$ harmonic order,  $J_{ql}(..)$ is the Bessel function of order $ql$, and {($\beta,\theta$)  represent the divergence, and azimuthal coordinate at the far-field, respectively. Our theoretical investigation mainly focuses on studying the divergence and phase profiles of different harmonic orders, therefore we set $C$ and $\tau^{j}$ to 1.
From the far-field intensity, we gain insight into the divergence profile of different harmonic orders and verify the upscaling law of OAM (as seen in the far-field phase profile, see Appendix \ref{A3}). Our numerical simulations reveal that all harmonics are emitted with similar divergence, as depicted in Figs.~\ref{PhaseAndIntensity}(a)-(c). This similarity is quantified by examining the radius of the maximum intensity ring, which remains consistent across different harmonic orders. It is also important to highlight that the radius of the maximum intensity bright ring for different harmonic orders is the same as that of the fundamental POV beam. This led to the fact that the POV beam structure is retained in the non-perturbative process of HHG. Additionally, an evaluation of the far-field phase profiles demonstrates that the OAM upscaling law ($l_{q}=ql$) is strictly adhered to. Specifically, the OAM of the 19$^{\mathrm{th}}$ ($\lambda_{19}=42.1$~nm), 21$^{\mathrm{st}}$ ($\lambda_{21}=38$~nm), and 23$^{\mathrm{rd}}$ ($\lambda_{23}=34.8$~nm) harmonics are found to be 19, 21, and 23, respectively, as illustrated in Figs.~\ref{PhaseAndIntensity}(d)-(f). It is also important to highlight that the law of OAM upscaling ($l_{q}=ql$) is strictly followed when a driving field composed of a single POV beam of any OAM interacts with the gas target. However, we expect this law to be modified in the case when a driving field comprises a mixture of POV beams with different OAMs. The far-field profile of the POV-based harmonics resembles that of the harmonics generated by an LG beam. However, employing the POV beam offers certain advantages, such as the ability to use beams with any value of OAM for HHG, which is not feasible with LG beams. This is due to the fact that the use of higher-order input LG beams generates XUV beams which are too weak to be detected and resolved as discussed in~\cite{zurch2012strong}. Another exciting possibility is the creation of attosecond pulses. Although the spatial intensity profile does not vary from harmonic to harmonic as it is locked, the OAM and the corresponding phase profile do. This will result in a complex spatially varying temporal attosecond pulse structure.

\begin{figure}[h!]
\includegraphics[width=0.8\textwidth, angle = -90]{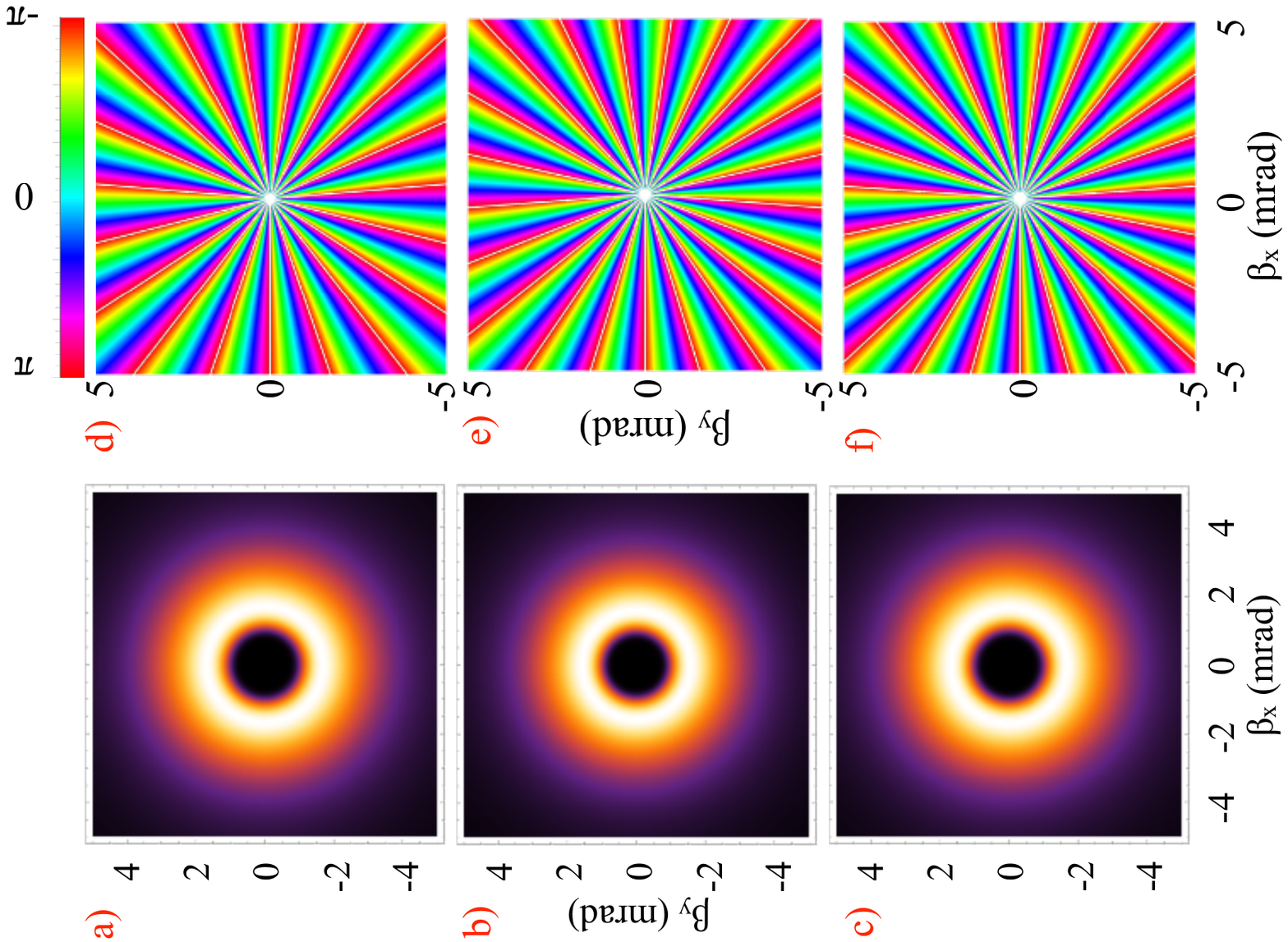}
\caption{\label{PhaseAndIntensity} 
Far-field intensity (a)-(c) and phase (d)-(f) profiles for various harmonics. (a) and (d) $19^{\mathrm{th}}$ ($\lambda_{19}=42.1$~nm), (b) and (e) $21^{\mathrm{st}}$ ($\lambda_{21}=38$~nm), (c) and (f) $23^{\mathrm{rd}}$ ($\lambda_{23}=34.8$~nm). The divergence of different harmonic orders is quantified by the radius of the maximum intensity ring, which is the same for all harmonic orders considered here. (d)-(f) illustrates the OAM upscaling law  ($l_{q}=ql$) (see the text for more details). }
\end{figure} 

\section{Conclusions and Outlook}
Our study investigated the non-perturbative interaction between a spatially structured linearly polarized POV beam and an atomic gas target. Near and far-field harmonic amplitudes and phases were computed using the TSM when the target was positioned at the beam focus. The results demonstrated that various harmonics exhibited similar divergence, adhering strictly to the upscaling OAM law ($l_{q}=ql$). Notably, perfect vortex harmonics, characterized by their divergence, were generated in the far-field. It is important to highlight that the spatially identical profiles of different harmonics will enable the generation of spatially locked attosecond pulses with non-trivial spatial variations easily. However, imperfections were observed between the fundamental and harmonic vortex profiles: the fundamental field resembled a spatially structured POV beam, while the harmonic vortex profile resembled that of an LG beam. This imperfection may stem from:(1) a poor phase-matching condition at the focus, attributable to interfering contributions from both short and long trajectories across different harmonic orders, (2) propagation of the harmonic vortices from the near to the far-field. This can be explained in the following way: At the near-field, the resulting harmonic vortices carry a similar spatial intensity profile as the driving POV beam. However, the propagation of the harmonic vortices to the far-field resulted in a spatial intensity profile which resembles to that of an LG beam profile. Similar propagation behavior has already been demonstrated for the POV beam propagating in free-space in the linear regime~\cite{Bikash24Elliptical}. Placing the gas jet behind the beam focus position could enhance the phase-matching condition. Moreover, the limitation of using low-order OAM beams, particularly LG beams, for HHG could be overcome by employing POV beams, which inherently possess an OAM-independent beam size and radial intensity distribution. Our research demonstrates the superiority of employing POV beams over LG beams, particularly in scenarios necessitating a high topological charge with a compact core size. Our findings pave the way towards the creation of bright, high-photon-flux, structured XUV coherent radiation sources. The TSM can also be applied directly to solid-state HHG as the last few tens of nm of the material contribute to HHG. These sources serve as crucial tools with diverse applications spanning numerous technological fields. For example, in stimulated emission depletion (STED) microscopy, modifying the selection rules in spectroscopy to allow the forbidden transitions, HHG spectroscopy, and finding new forms of dichroism.

\begin{acknowledgments}

We acknowledge financial support from the Guangdong Province Science and Technology Major Project (Future functional materials under extreme conditions - 2021B0301030005) and the Guangdong Natural Science Foundation (General Program project No. 2023A1515010871).

\end{acknowledgments}

\appendix

\section{Perfect Optical Vortex beams formulae} \label{A1}

The spatial complex field amplitude of a Perfect Optical Vortex (POV) beam is given by \cite{vaity2015spatial}: 
\begin{equation}
    E(r,\theta,z=0)=E_0 e^{-\frac{r^2+R^2}{\omega_0^2}}e^{il\theta}I_l\Bigg(\frac{2rR}{\omega_0^2}\Bigg),
\end{equation}
where $E_0$, $R$, $\omega_0$, and $l$ represent the constant field amplitude, radius, half-ring width, and OAM (or, TC), respectively. $I_l(..)$ denotes the modified Bessel function of the first kind. The phase term $\exp(il\theta)$ represents an arbitrary helical phase. The transverse intensity profile of a POV beam of TCs $1$, $2$, $5$, and $10$ are shown in Fig.~\ref{pov}. Here, we can clearly observe that the intensity distribution and core size of the POV beams are independent of the TCs. 

\begin{figure}[h!]
\includegraphics[width=0.75\textwidth]{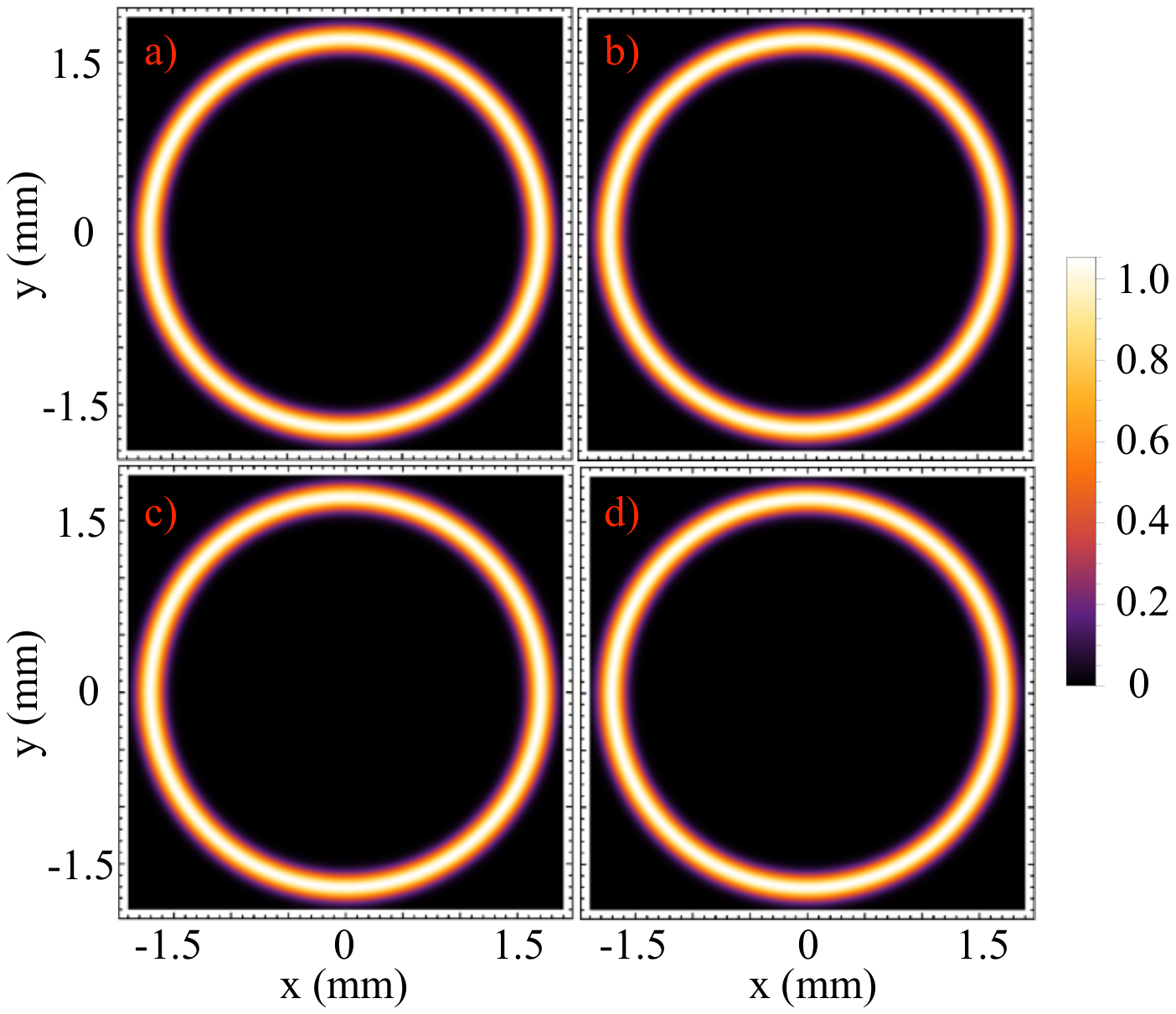}
\caption{The transverse intensity profile of the POV beam of TC (a) 1, (b) 2, (c) 5, and (d) 10.}
\label{pov}
\end{figure} 

The complex field amplitude of a linearly polarized spatio-temporal POV beam reads: 

\begin{eqnarray}
    E(r,\theta,z=0,t)&=&E_0 e^{-\frac{r^2+R^2}{\omega_0^2}}e^{il\theta}I_l\Bigg(\frac{2rR}{\omega_0^2}\Bigg)\text{sin}^2\Bigg(\frac{\pi t}{n T}\Bigg)\text{cos}\Big( \omega t + \varphi_{CE} \Big)\label{POVCA}.
\end{eqnarray}
where $\omega$, $\varphi_{CE}$, $n$ and $T=2\pi/\omega$ represent the central frequency of the driving laser field, the carrier-envelope phase, the total number of cycles and the laser period, respectively. To remove the complexity of solving the far-field integral involving the modified-Bessel function $I_l$, we approximate it as an exponentially growing function. This is justified under the condition $R\gg\omega_0$. Thus, 
\begin{equation}
    I_l\Bigg(\frac{2rR}{\omega_0^2}\Bigg)\approx e^{\frac{2rR}{\omega_0^2}}.
\end{equation}
Thus, Eq.~(\ref{POVCA}) takes the form: 
\begin{eqnarray}
    E(r,\theta,t)&=&E_0 e^{-\frac{(r-R)^2}{\omega_0^2}}e^{il\theta} \text{sin}^2\Bigg(\frac{\pi t}{n T}\Bigg)\text{cos}\Big( \omega t + \varphi_{CE} \Big)\label{POVCA2}.
\end{eqnarray}
Equation~(\ref{POVCA2}) was used to calculate the HHG spectra under the strong field approximation (SFA). In our simulations, $I_0=1.57\times 10^{14}$ W/cm$^2$, $\omega_0=150$ $\mu$m, $R=$ 1.71 mm, $n=64$ cycles, $\lambda=800$ nm, $\varphi_{CE}=0$, $T=2.67$ fs, and $l=1$. 

To calculate the near and far-field complex amplitudes and phases of different harmonic orders, we employ a relatively simple yet efficient and accurate model: the Thin-Slab Model (TSM) \cite{usev}. In this approach, the actual target is replaced with a thin 2D slab positioned perpendicular to the propagation axis of the driving laser beam. The beam intensity is maximal at the ring of the POV beam. Consequently, we can disregard the longitudinal phase matching and focus solely on the transverse phase matching. Additionally, it is worth noting that since we position the slab at the beam's focal point ($z=0$), any phase contributions stemming from the curved wavefront (focal phase) and focusing (Gouy phase) of the beam can be easily avoided. Only the helical and dipole phases are taken into consideration.

For the calculation of the HHG, we consider first the contributions arising from the single-atom interaction (dipole phase) and the helical phase. The beam profile of the driving field in the thin slab placed at $z=0$ is giving by:
\begin{equation}
    A(r',\theta ') = U(r')e^{i \varphi_{t}(\theta ')},
\end{equation}
where the total phase $\varphi_t=l\theta '$. Also, we need to consider $k_0r^2/(2R(z))=0$ as $R(z)\rightarrow \infty$ at $z=0$. Again, the Gouy phase also vanishes at $z=0$. In this model, we consider that the high-order harmonics are emitted at the thin slab. We can calculate the contribution of the $j^{\mathrm{th}}$ quantum path to the $q^{\mathrm{th}}$ harmonic field at a slab located at $z=0$ as:

\begin{eqnarray}
    A_q^{j(\text{near})}(r',\theta ',z=0)&=&\Bigg(\frac{C}{\tau^j}\Bigg)^{3/2}\abs{U(r')}^p e^{iql\theta '+ i\alpha _q^j\abs{U(r',\theta ')}^2}. 
\end{eqnarray}
Here, $A_q^{j}(r',\theta ',z=0)$ is the near-field amplitude of the $q^{\mathrm{th}}$ harmonic order, $C$ ($=1$ in our simulation) is a constant, $\tau^{j}$ is the excursion time associated with the $j^{\mathrm{th}}$ quantum path, $\abs{U(r')}$ corresponds to the amplitude of the fundamental POV field and $e^{iql\theta '+ i\alpha _q^j\abs{U(r',\theta ')}^2}$ corresponds to the phase term which accommodates both helical and dipole phase of the $q^{th}$ order harmonic . $p$ is a scaling factor which determines how every harmonic amplitude scales with the fundamental field amplitude. We will calculate $p$ for the harmonics considered in our simulations using the SFA.

\section{Strong-field approximation}\label{A2}
The time-dependent Schrödinger equation (TDSE), which describes the interaction of an atom with an intense laser field under the single active electron approximation, within the dipole approximation, and in length gauge, reads: 
\begin{equation}
    i\frac{\partial}{\partial t} |\Psi(\mathbf{r},t)\rangle=\Big[ -\frac{1}{2}\nabla^2+V(r)-\mathbf{r}\cdot\mathbf{E}(t)\Big]|\Psi(\mathbf{r},t)\rangle,
\end{equation}
where $H_0=-\nabla^2/2+V(r)$ and $H_I=-\mathbf{r}\cdot\mathbf{E}(t)$ are the field-free and interaction Hamiltonian respectively. The Lewenstein model \cite{maciejSFA} is used to solve this TDSE with the following assumptions: (i) the laser dressing of the excited bound states is neglected ($I_p \gg \hbar \omega$), (ii) the ground state depletion is ignored ($I \ll I_{sat}$) and (iii) the effect of the Coulomb potential on the ionized electron in the continuum is neglected ($I_p \ll 2U_p$). Thus, the freed electron evolves under the sole influence of the oscillating laser field. Following all these assumptions, the TDSE can be solved analytically. The ansatz used to solve this equation reads as follows:
\begin{equation}
    |\Psi(r,t)\rangle=e^{-iI_p t}\Big( a(t)|g\rangle + \int d\mathbf{k} b(\mathbf{k},t)|\mathbf{k}\rangle \Big),
\end{equation}
where $a(t)$ is the time-dependent ground state amplitude which is set to 1 (for weak ionization) and $b(\mathbf{k},t)$ is the continuum states amplitude, which depends on the kinetic momentum and the laser field. $|g\rangle$ and $|k\rangle$ are the ground and continuum eigenstates of the field-free Hamiltonian, respectively. The time-dependent dipole moment can then be written as:
\begin{eqnarray}
    D(t)&=&i\int_{-\infty}^{t}dt'\int d^{3}\mathbf{p}\,d^*(\mathbf{p}+\mathbf{A}(t)) \nonumber \\
    &&e^{-i\int_{t'}^t (S(\mathbf{p},t,t')) dt'' } \mathbf{E}(t')\, d(\mathbf{p}+\mathbf{A}(t')).
\end{eqnarray}
Thus, the HHG yield can be found as
\begin{equation}
    P(\omega)=\Bigg| \int_{-\infty}^{\infty} D(t)e^{-i\omega t} dt \Bigg|^2.
\end{equation}
This last equation allows us to compute the single atom response when an ultrashort laser field interacts with an atom. The time-dependent dipole moment $D(t)$ can be understood in terms of the three-step model, i.e., (a) tunnel ionization, (b) propagation of the freed electron in the laser field, and (c) recombination with the parent ion core, when the laser field direction is reversed. Spectra computed using SFA for different laser intensities are presented in Fig.~\ref{hhgyield}. Note that the full blown SFA does not distinguish long and short trajectories and therefore contains both.
\begin{figure}[h!]
\includegraphics[width=.8\textwidth]{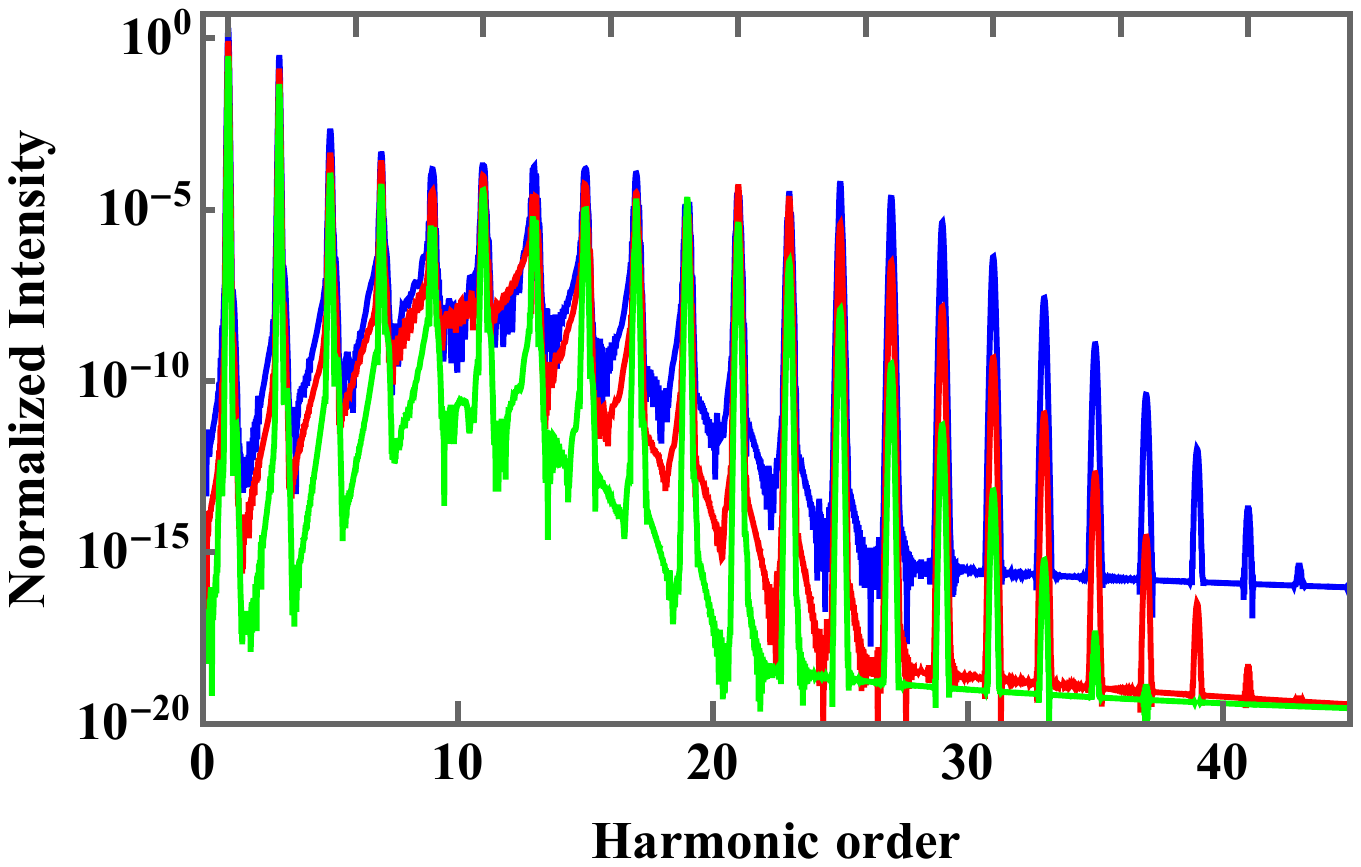}
\caption{HHG spectrum computed using the SFA for three different laser intensities  $0.875\times10^{14}$ W/cm$^{2}$ (red curve), $1.177\times10^{14}$ W/cm$^{2}$ (green curve), $1.524\times10^{14}$ W/cm$^{2}$ (blue curve). }
\label{hhgyield}
\end{figure}

\subsection{Far field calculations} \label{A3}
The fundamental parameter for the calculation of the far-field amplitudes is the scaling parameter $p$. For the perturbative (non-perturbative) case, $U_q\propto U^q_{f}$ with $p=q$ ($p<q$). Here, $U_{q}$, and $U_{f}$ correspond to the amplitudes of the $q^{\mathrm{th}}$ harmonic order and the fundamental field, respectively. For the computation of the near-field amplitude of different harmonic orders, we used the SFA and extract the field amplitudes at the slab. We employed three different field intensities of the driving field, namely $I_1=0.875\times 10^{14}$ W/cm$^2$, $I_2=1.177\times 10^{14}$ W/cm$^2$ and $I_3=1.542\times 10^{14}$ W/cm$^2$. By plotting $\log(U_q)$ vs. $\log(U_{f})$, we extract the values of $p$ for different harmonic orders. Thus $p$ becomes 3.9, 3.4, and 3.6 for harmonic orders $19^{\mathrm{th}}$, $21^{\mathrm{st}}$, and $23^{\mathrm{rd}}$, respectively. Note that, different values of the scaling parameter, $p$, were calculated numerically following the scheme presented in~\cite{usev}. Additionally, it is important to highlight that the numerical integration was performed in 3D without the use of the saddle point approximation.

For computing the far-field harmonic amplitude, we use the Fraunhofer-diffraction integral as follows: 
\begin{eqnarray}
    A_q^{j(\text{far})}(\beta, \theta)&\propto& \Bigg(\frac{C}{\tau^j}\Bigg)^{3/2}\int_0^{\infty} \int_0^{2\pi}r' dr' d\theta ' A_q^{j(\text{near})}(r')  e^{-i\frac{2\pi r'}{\lambda_q} \text{tan}(\beta)\text{cos}(\theta-\theta')}\nonumber \\
    &\propto& \Bigg(\frac{C}{\tau^j}\Bigg)^{3/2}\int_0^{\infty} \int_0^{2\pi}r' dr' d\theta' \abs{U(r')}^{p}e^{iql\theta '}  e^{i\alpha_q^j \abs{U(r',\theta ')}^2}e^{-i\frac{2\pi r'}{\lambda_q} \text{tan}(\beta)\text{cos}(\theta-\theta')}. 
\end{eqnarray}

Here $U(r',\theta ')$ is the fundamental POV beam field, and ($\beta$, $\theta$) are the far-field coordinates representing the divergence, and the azimuthal coordinate respectively. By solving the angular integral we can write: 
\begin{eqnarray}
    A_q^{j(\text{far})}(\beta, \theta)&\propto& \Bigg(\frac{C}{\tau^j}\Bigg)^{3/2}e^{iql\theta}i^{ql}  \int_0^\infty r' dr'\abs{U(r')}^p J_{ql}\Bigg( \frac{2\pi\beta r'}{\lambda_q}  \Bigg)e^{i\alpha_q^j \abs{U(r')}^2}.
\end{eqnarray}

Notice that the dipole phase term is taken into account in the single atom response calculation from the SFA. The amplitude part: 
\begin{equation}
    \abs{U(r')}^p=E_0^p e^{-p\frac{(r'-R)^2}{\omega_0^{2}}}
\end{equation}
allow us to write: 
\begin{eqnarray}
A_q^{j(\text{far})}(\beta, \theta)&\propto& \Bigg(\frac{C}{\tau^j}\Bigg)^{3/2}e^{iql\theta}i^{ql}  \int_0^\infty r' dr'E_0^p e^{-p\frac{(r'-R)^2}{\omega_0^{2}}} J_{ql}\Bigg( \frac{2\pi\beta r'}{\lambda_q} \Bigg)e^{i\alpha_q^j \abs{U(r')}^2}.
\end{eqnarray}
This integral can be solved analytically by binomially expanding the exponential part of the amplitude of the integrand.  

Thus, the intensity of far-field harmonics can be calculated from:
\begin{eqnarray}
\abs{A_q^{j(\text{far})}(\beta, \theta)}^2&\propto& \Bigg[\Bigg(\frac{C}{\tau^j}\Bigg)^{3/2} e^{iql\theta}i^{ql} \Bigg(\frac{2\pi\beta}{\lambda_q}\Bigg)^{ql} E_0^p e^{-\frac{pR^2}{\omega_0^2}} \sum_{n=0}^\infty \Bigg(\frac{2pR}{\omega_0^2}\Bigg)^n\frac{1}{n!}\frac{\Gamma \Bigg( \frac{ql+n+2}{2}\Bigg)}{2^{ql+1}\Big(\frac{p}{\omega_0^2}\Big)^{\frac{n+ql+2}{2}}\Gamma(ql+1)} \nonumber \\
&\times& _1F_1\Bigg( \frac{n+ql+2}{2}; ql+1;-\frac{\pi^2\beta^2 \omega_0^2}{p\lambda_q^2} \Bigg)\Bigg]^2
\end{eqnarray}
where $\lambda_q=\lambda/q$. 
The intensity, $I(\beta)=\abs{A_q^{j(\text{far})}(\beta, \theta)}^2$, for $n=0$ now takes compact the form: 
\begin{equation}
    I(\beta)\propto \Bigg(\frac{C}{\tau^j}\Bigg)^{3}I_0^p e^{\frac{-2p R^2}{\omega_0^2}}\frac{\Big(\frac{2\pi\beta}{\lambda_q}\Big)^{2ql} \Big[\Gamma \Big( \frac{ql+2}{2}\Big)\Big]^2}{2^{2ql+2}\Big(\frac{p}{\omega_0^2}\Big)^{ql+2}\Big[\Gamma(ql+1)\Big]^2} \Bigg[{_1}F_1\Bigg( \frac{ql+2}{2}; ql+1;-\frac{\pi^2\beta^2 \omega_0^2}{p\lambda_q^2} \Bigg)\Bigg]^2.
\end{equation}

\section{Calculation of the divergence for LG and POV beams}
\textit{Case 1:}
The complex spatial field amplitude of the LG beam can be written as: 

\begin{equation}
    E(r,\theta,z)=E_0\frac{\omega_0}{\omega(z)}\Bigg( \frac{\sqrt{2}r}{\omega(z)}\Bigg)^le^{-\frac{r^2}{\omega(z)^2}} e^{il\theta}e^{\frac{ikr^2}{2R(z)}}e^{i\varphi_G(z)},
\end{equation}
where $E_0$, $\omega_0$ and $l$ are the constant field amplitude, beam waist, and topological charge of the LG beam, respectively.
The beam size, $\omega(z)$, the radius of curvature, $R(z)$, and the Gouy phase, $\varphi_G(z)$, of the LG beam are defined as follows: 
\begin{eqnarray}
    \omega(z) &=& \omega_0 \sqrt{1+\frac{z^2}{z^2_R}}, \nonumber \\
    R(z) &=& z\Bigg(1+\frac{z^2_R}{z^2}\Bigg),\nonumber\\
    \varphi_G(z) &=& -(2p+l+1)\text{tan}^{-1}\Bigg(\frac{z}{z_R}\Bigg).
\end{eqnarray}
Here, $p$, and $l$ denote the radial and azimuthal (or, topological charge) indices of the LG beam respectively, and $z_{R}$ corresponds to the Rayleigh range ($=\frac{1}{2} k\omega_{0}^2$).

Now, the intensity of the LG beam can be approximated to 
\begin{equation}
    I(r,z) \approx  \frac{1}{\omega^2(z)}\Bigg( \frac{\sqrt{2}r}{\omega(z)}\Bigg)^{2l} e^{-\frac{2r^2}{\omega(z)^2}}.
\end{equation}
In the case of an LG beam, the intensity is maximum on a ring. This means, the radius of the maximum intensity can be obtained by using the following conditions, 
\begin{eqnarray}
    \frac{\partial I(r,z)}{\partial r}&=&0 \nonumber \\ 
    \text{or } \frac{1}{\omega^2(z)} \frac{\partial}{\partial r}\Bigg[\Bigg( \frac{\sqrt{2}r}{\omega(z)}\Bigg)^{2l} e^{-\frac{2r^2}{\omega(z)^2}} \Bigg]&=&0. \nonumber
\end{eqnarray}

\begin{eqnarray}
    \frac{1}{\omega^2(z)} \Bigg( \frac{\sqrt{2}r}{\omega(z)}\Bigg)^{2l} e^{-\frac{2r^2}{\omega(z)^2}}\Bigg( \frac{2l}{r}-\frac{4r}{\omega^2(z)} \Bigg)&=&0 \nonumber \\
    \text{or } \frac{2l}{r_{\mathrm{max}}}-\frac{4r_{\mathrm{max}}}{\omega^2(z)}&=&0,
\end{eqnarray}
with $r_{\mathrm{max}}=(l/2)^{1/2}\omega(z)$. It is found that the radius of the maximum intensity for an LG beam depends strictly on the topological charge of the beam.

We can now calculate the divergence of the LG beam, which is given by:
\begin{equation}
    \beta_{\mathrm{div}}=\lim_{z\rightarrow \infty}\text{tan}^{-1}\Bigg (\frac{\partial r_{\mathrm{max}}}{\partial z} \Bigg) \approx \sqrt{\frac{l}{2}}\frac{\lambda}{\pi \omega_0} 
\end{equation}
It is evident from the above equation that the divergence of the LG beam is dependent on the topological charge, and the Gaussian beam waist at the focus.

\textit{Case 2}: For the POV beam, the spatial complex field amplitude can be written as: 
\begin{equation}
    E(r,\theta,z)=E_0\frac{\omega_0}{\omega(z)}
    e^{-\frac{1}{\omega(z)^2}\Big(r^2-\frac{R^2z^2}{z_R^2}\Big)}e^{i\psi_G(z)}e^{il\theta}
    e^{\frac{ikr^2}{2R(z)}(r^2+R^2)}I_l\Bigg( \frac{2rR\,e^{i\psi_G(z)}}{\omega_0\omega(z)}\Bigg),
\end{equation}
where $E_0$, $\omega_0$, $R$ and $l$ are the constant field amplitude, the Gaussian beam waist, the radius and the topological charge of the POV beam, respectively.
The Gaussian beam size, $\omega(z)$, the radius, $R(z)$, and the topological charge, $\psi_G(z)$, of the POV beam are defined as follows: 
\begin{eqnarray}
    \omega(z) &=& \omega_0 \sqrt{1+\frac{z^2}{z^2_R}}, \nonumber \\
    R(z) &=& z\Bigg(1+\frac{z_R^2}{z^2}\Bigg),\nonumber\\
    \psi_G(z) &=& -\text{tan}^{-1}\Bigg(\frac{z}{z_R}\Bigg).
\end{eqnarray}
If the radius of the POV beam is much higher than that of its half-ring width i.e., $R\gg\omega_0$, we can write
\begin{equation}
    I_l\Bigg( \frac{2rRe^{i\psi_G(z)}}{\omega_0\omega(z)}\Bigg) \approx e^{\frac{2rRe^{i\psi_G(z)}}{\omega_0\omega(z)}}.
\end{equation}
For this case, the complex field amplitude of the POV beam can be rewritten as,
\begin{equation}
    E(r,\theta,z)=E_0\frac{\omega_0}{\omega(z)}
    e^{-\frac{1}{\omega(z)^2}\Big(r^2-\frac{R^2z^2}{z_R^2}\Big)}e^{i\psi_G(z)}e^{il\theta}
    e^{\frac{ikr^2}{2R(z)}(r^2+R^2)}e^{\frac{2rR\psi_G(z)}{\omega_0\omega(z)}}.
\end{equation}
Now, the intensity of the POV beam reads:
\begin{equation}
    I(r,z) \approx\frac{1}{\omega^2(z)}
    e^{-\frac{2r^2}{\omega(z)^2}}
    e^{\frac{4rR}{\omega_0\omega(z)}\text{cos}(\psi_G(z))}.
\end{equation}

In the case of a POV beam, the intensity is maximum on a ring. The radius of the maximum intensity can be obtained by using: 

\begin{eqnarray}
    \frac{\partial I(r,z)]}{\partial r}&=&0 \nonumber \\ 
    \text{or } \frac{1}{\omega^2(z)} \frac{\partial}{\partial r}\Bigg(   e^{-\frac{2r^2}{\omega(z)^2}}
    e^{\frac{4rR}{\omega_0\omega(z)}\text{cos}(\psi_G(z))} \Bigg)&=&0 \nonumber
\end{eqnarray}

\begin{eqnarray}
    \frac{1}{\omega^2(z)}   e^{-\frac{2r^2}{\omega(z)^2}}
    e^{\frac{4rR}{\omega_0\omega(z)}\text{cos}(\psi_G(z))} \Bigg(-\frac{4r}{\omega^2(z)}+\frac{4R  \text{cos}(\psi_G(z))}{\omega_0 \omega(z)} \Bigg)&=&0 \nonumber\\
    \text{or }-\frac{4r_{\mathrm{max}}}{\omega^2(z)}+\frac{4R  \text{cos}(\psi_G(z))}{\omega_0 \omega(z) }=0, \nonumber
    \end{eqnarray}
where $r_{\mathrm{max}} = R \omega(z) \text{cos}(\psi_G(z))/\omega_0$. It is found that the radius of the maximum intensity i.e., the divergence of the POV beam is independent of the topological charge of the beam. Therefore, the POV beam follows the same divergence law as of a fundamental Gaussian beam~\cite{vaity2015spatial}. For larger propagation distances i.e., at $z \gg z_{R}$, the beam size scales linearly with the propagation distance, $z$. However, $\cos(\psi_{G}(z))$ decreases significantly as $\psi_{G}(z)$ approaches $\frac{\pi}{2}$, for $z \gg z_{R}$. Hence, the radius of the maximum intensity of the ring decreases~\cite{Bikash24Elliptical} as the beam propagates to larger distances.


\bibliography{literatur}


\end{document}